\documentclass[twocolumn]{aastex631} 

\usepackage{amsmath,cancel}

\newcommand{\be}{\begin{eqnarray}}
\newcommand{\ee}{\end{eqnarray}}
\newcommand{\bi}{\begin{itemize}}
\newcommand{\ei}{\end{itemize}}

\newcommand{\R}{{\cal R}}
  
\newcommand{\bld}[1]{\mbox{\boldmath$#1$\unboldmath}}
\newcommand{\dd}{{\partial}}

\begin{document}

\title{Enceladus's Tidal Heating}

\correspondingauthor{Yoram Lithwick}
\email{y-lithwick@northwestern.edu}
\author[0000-0003-4450-0528]{Yoram Lithwick}
\affiliation{Department of Physics \& Astronomy, Northwestern University, Evanston, IL 60202, USA}
\affiliation{Center for Interdisciplinary Exploration \& Research in Astrophysics (CIERA), Evanston, IL 60202, USA}

\begin{abstract}
Saturn raises a time-dependent tide on its small moon Enceladus, due to the eccentricity
of the orbit.
As shown in a companion paper  (Goldreich et al. 2025), the resulting
tidal heating drives
 Enceladus into
  a limit cycle, 
in which its  eccentricity and shell thickness vary in tandem, on a timescale of $\sim 10$Myr. 
The limit cycle  explains
 a variety of observed phenomena on Enceladus, including its
large luminosity
 and cracked ice shell.
Here we derive the tidal heating rate needed for that study, 
starting from a simple first-principles derivation of  Enceladus's tidal response.
Enceladus is  comprised of three layers: a rocky core, 
an outer ice shell, and an ocean sandwiched in between. 
Tides force the  shell to librate and distort, which
generates heat. 
We calculate the libration amplitude and tidal heating rate 
 by minimizing
the sum of elastic and gravitational energies. The final expressions are
 analytic, and account for  the finite hardness of the shell, 
and for   resonant libration. 
Although we specialize to Enceladus, 
our approach may be  extended to other bodies
 that have a similar three layer structure,
 such as Europa and Titan.

\end{abstract}

\section{Introduction}

Enceladus exhibits a range of surprising behaviors, including water spraying from its surface, 
a librating ice shell floating on a global ocean, and an abundance heat escaping from its surface. 
In a companion paper (Goldreich et al. 2025), we show that these phenomena can 
be understood as a consequence of Enceladus being in a limit cycle. 
One of the main ingredients of the limit cycle is the tidal heating rate of Enceladus ($H$), 
including how $H$ depends on the moon's eccentricity and shell thickness. 
The main goal of this paper is  to calculate the heating rate. 
For an introduction to the phenomenology of Enceladus, see the references
cited
in Goldreich et al. 

\cite{2010Icar..209..631G} and \cite{2013Icar..226..299V} consider the effect of tides on moons such as Enceladus
that have an ice shell atop an ocean.  Tides are driven by the moon's eccentric
orbit about the planet, and the tides in turn drive 
  forced librations  of  the moon's shell. 
\cite{2019Icar..332...66B} and \cite{2022Icar..37314769S} have previously 
calculated the  energy dissipation that is caused by the 
distortions of the ice shell as it undergoes its forced librations. 
Our derivation largely reproduces their results, but there are a few differences:
\bi
	\item We provide a simple and  first-principles derivation that
	is based on minimizing the sum of gravitational and elastic energies.  
	\item Our final expression is analytic (equation \ref{eq:heatfin}). 
	It  accounts for what happens whether the shell is hard or soft, 
	as well as for what happens near resonant libration.
\ei

\section{Saturn's Potential}

We repeat here the well-known expressions for the planet's potential as experienced by an eccentric moon
  \citep{1999ssd..book.....M},  because it   will help us introduce  our notation.
  We work in a reference frame centered on Enceladus, and rotating at the frequency of Enceladus's
mean motion around Saturn ($n$). We adopt spherical coordinates ($r,\theta,\phi$),  where 
Saturn's guiding center is at
 $\phi=0$.
        We decompose
        Saturn's quadrupolar potential at the surface of the undistorted Enceladus (i.e., at $r=R$, where
        $R$ is Enceladus's undistorted radius)  in 
        spherical harmonics:
       \be
       	       V(\theta,\phi,t) =  V_0 Y_{\ell,0} +  V_+ Y_{\ell,+} + V_-Y_{\ell,-}
	       	       \label{eq:vexp}
       \ee
	   where $\ell\equiv 2$ throughout this paper, and the  $Y_{\ell,m}$'s are  real spherical harmonics,
	   \be
	   Y_{\ell,0}&=& {1\over 4}\sqrt{5\over\pi}\left(3\cos^2\theta-1  \right)
	   \\
	   Y_{\ell,+}&=&  {1\over 4}\sqrt{15\over\pi}\sin^2\theta\cos 2\phi
	   \\
	   Y_{\ell,-}&=& {1\over 4}\sqrt{15\over\pi}\sin^2\theta\sin 2\phi
	   \ ,
	   \ee
	   which are orthonormal.  We label the three $m$'s with $m = \{ 0, +, -\}$
	   rather than  $\{0, 2, -2 \}$ to avoid confusion with other subscripts
	   to be defined later. 
	    The three  $V_m$'s
	   are time-dependent because the orbit is eccentric. 
	   For the remainder of this subsection only, we  collect the $V_m$'s  into a column vector
	   \be
	   \bld{V}  = 
	   \begin{bmatrix} V_0 \\ V_+ \\ V_{-}  \end{bmatrix}  \ .
	   \ee
	   The potential may  be decomposed into static and epicyclic components:
	  	  \be
	    \bld{V} = \bld{\bar{V}} + \bld{V'}(t) \ , \label{eq:vdecomp}
	   \ee
	   where the static component is
	   \be
	   \bld{\bar{V}} &=& n^2R^2\sqrt{\pi\over 5}\left( \begin{bmatrix} 1 \\ -\sqrt{3} \\ 0  \end{bmatrix} 
	   + \begin{bmatrix} 2/3 \\ 0 \\ 0  \end{bmatrix}   \right) \label{eq:vbar}
	   \ee
	   and the epicyclic is
	   \be
	   \bld{V'}(t) &=& n^2R^2\sqrt{\pi\over 5}\left(  \begin{bmatrix} 3 \\ -3\sqrt{3} \\ 0  \end{bmatrix}e\cos n t - 
	   \begin{bmatrix} 0 \\ 0 \\ 2\sqrt{3}  \end{bmatrix}\gamma_S \label{eq:vprime}
	    \right)
	    \ \  \ \ 
	   \ee
	   \citep{1999ssd..book.....M}.
	   The first column vector in $\bld{\bar{V}}$  is from the  tidal potential and the second is from the centrifugal.
	   And the first in $\bld{V'}$ is from expanding the static tidal potential to linear
	   order in Saturn's radial epicyclic excursion ($-e\cos nt$), and the second
	   is from expanding in its azimuthal excursion
	   \be
	   \gamma_S \equiv 2e\sin nt \ . \label{eq:gsdef}
	   \ee
\begin{figure}[h]
\centering
    \includegraphics[width=.53\textwidth]{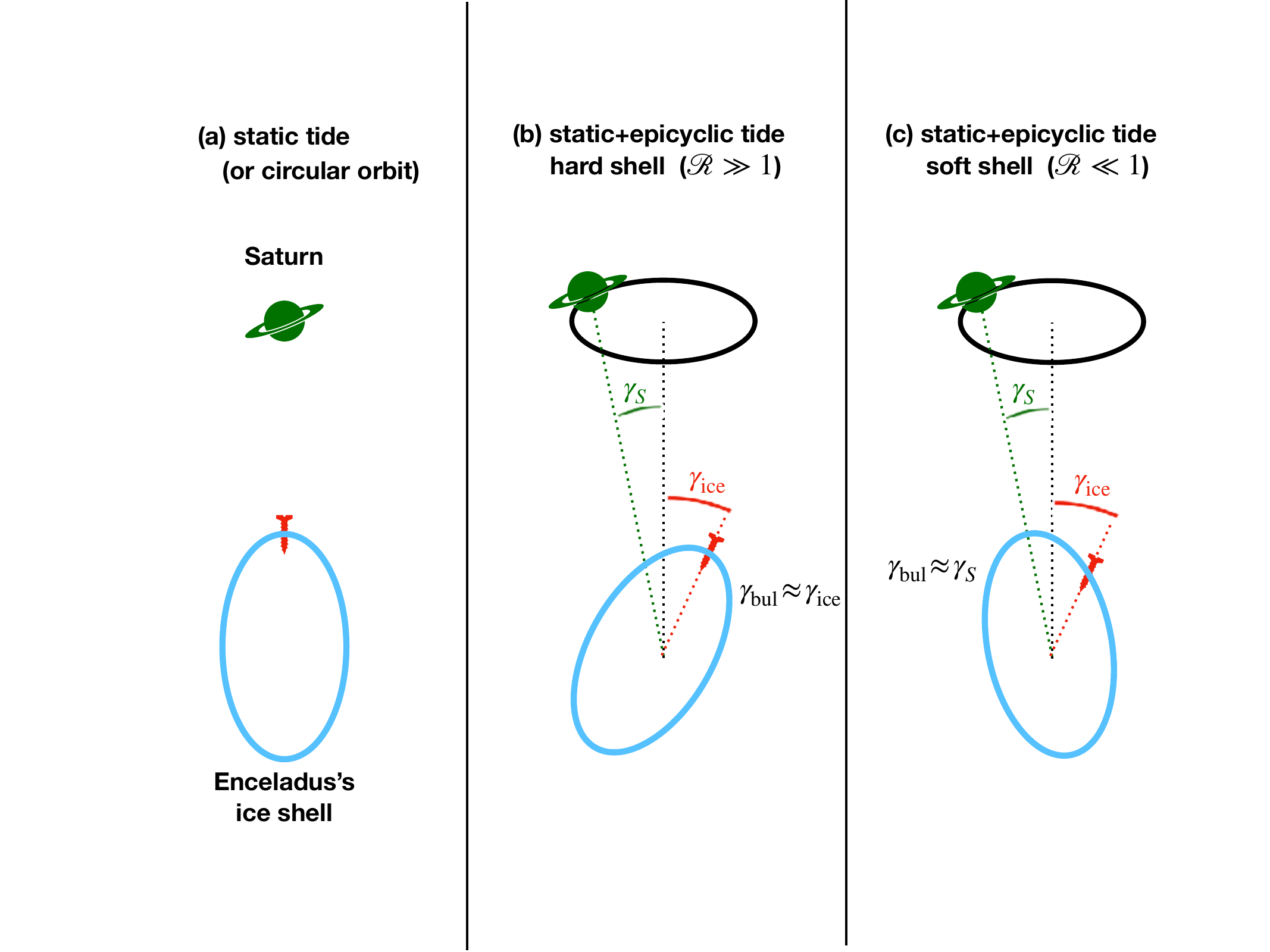}
\caption{{\bf Orientation of Tidal Bulge:} Panel (a) displays the bulge of the ice shell
under the action of the static tide, and the other panels  display the bulge under the
action of the total tide, depending on whether the shell is hard (panel b) or soft (panel c). 
For  given values of  $\gamma_S$ and $\gamma_{\rm ice}$ 
 (angles of Saturn in its epicycle and  of the shell, respectively), 
 the resulting orientation of the bulge ($\gamma_{\rm bul}$)
depends on the shell's hardness. When it is hard, the shell hardly changes its shape relative
to $\gamma_{\rm ice}$.  And when it is soft, the bulge nearly points at Saturn, whatever the value of 
$\gamma_{\rm ice}$. 
The hardness parameter ${\cal R}$ is defined in equation (\ref{eq:hardness}).
}
\label{fig:cartoon}
\end{figure}

       \section{Outline of Tidal Calculation}

We now work out  Enceladus's reaction to Saturn's potential. 
An important consideration is 
whether the shell is hard or soft \citep{2010Icar..209..631G}.  For a hard shell, the orientation
of the tidal bulge nearly tracks the orientation of the ice shell. Whereas for
a soft shell, the tidal bulge nearly points at Saturn
(Fig. \ref{fig:cartoon}). 
With Enceladus's current parameters, as listed in 
 Appendix \ref{sec:numbers}, the  shell is hard, 
 as will be shown  below. 
 But the thickness of Enceladus's shell varies over
 the course of a limit cycle.  If it is thin enough ($\lesssim 0.83$km), 
 it becomes soft.

An outline of 
 the tidal calculation   is as follows:
\bi
  \item In \S \ref{sec:toy}, we consider  a simplified two-layer model for Enceladus, 
  ignoring the core.  This produces nearly the same result as the full three-layer
  model considered later, aside from a few order-unity coefficients.  But it is considerably simpler algebraically, and
  so brings out the key physics, including the role of the hardness parameter ${\cal R}$. 
  \bi  \item  In \S  \ref{eq:statictide}, we derive   the static tide, 
  for which 
 the elastic strength of the shell may be ignored.
        \item In \S \ref{sec:etide}, we derive  the epicyclic tide, for which the elastic strength must 
        be included.
          Our calculation proceeds by first fixing $\gamma_S$
        and $\gamma_{\rm ice}$, as labelled in Figure \ref{fig:cartoon}, and then finding
        the resulting 
        $\gamma_{\rm bul}$ by minimizing the sum of elastic and gravitational energies.
         In addition to $\gamma_{\rm bul}$, 
        which comes from the $m=-$ component of the tide, we 
         solve for the $m=0$ and $+$ components, which will be needed for the heating rate. 
        We then use $\gamma_{\rm bul}$ to calculate the torque.  The torque is needed to solve
        for $\gamma_{\rm ice}(t)$. But we defer solving for $\gamma_{\rm ice}$ to the next section.

   \ei
   
   \item  In \S\S \ref{sec:threestat}--\ref{sec:threeepi}, we repeat the previous calculation, but now for the full three-layer
   model. 
        And in \S \ref{subsec:gamice},  we use the torque to solve for $\gamma_{\rm ice}(t)$.
       \item In \S \ref{sec:heatingrate}, we use $\gamma_{\rm ice}(t)$, as well as the $m=0$ and $+$ components of the tide, to obtain the tidal heating rate.
\ei

We make a number of approximations throughout this paper, aiming for an accuracy of $\sim 20\%$. 
We assume that the shell is very  thin and massless,\footnote{Accounting for the shell's mass
introduces a correction that is of order $\epsilon \equiv gR\rho_w/\mu$
\citep{2010Icar..209..631G}. 
For Enceladus, $\epsilon = 0.007$, and so we ignore it. 
But 
for Titan and Europa, 
$\epsilon\sim 1$, and the correction must be included, as has been done
in 
 \cite{2010Icar..209..631G}. 
 \label{foot:eps}
} and that the water is incompressible. We also ignore the 10\% difference in densities between 
water and ice.

	   \section{Two Layer Model}
	   \label{sec:toy}
	   
	    In this section we  model Enceladus as being entirely liquid water, aside from a thin ice shell
	     on top.  We include a core in the next section. 
	   The  response to Saturn's potential is completely specified by the radial displacement of the ice shell at Enceladus's surface,  $\xi$.  
	   We decompose $\xi$ in the same manner as the potential:
	   \be
	   \xi(\theta,\phi,t) = \xi_0Y_{\ell,0}+\xi_+Y_{\ell,+}+\xi_-Y_{\ell,-} \label{eq:xiexp}
	   \ee
	    Geometrically,  $\xi_0$ quantifies  flattening  along the polar axis; $\xi_+$
	   quantifies the amplitude of the tidal bulge
	   that is aligned with Saturn's guiding center; and $\xi_-$ quantifies the reorientation of the bulge  away from the guiding center.

	   \subsection{Static Tide}
	   \label{eq:statictide}
	   
	   We  determine the response of the ice shell to the static tide, $\bar{V}_m$ (Fig. \ref{fig:cartoon}a). 
	   Since the static tide is persistent,  elastic stresses in the shell have time
	   to relax away, and the shape of the shell is  determined by gravity alone. Quantitatively, 
	   we find  $\xi_m$  by minimizing the gravitational energy. 
	   For each of the three $m$'s, 
	    the gravitational energy relative to the unperturbed (spherical) state only depends on  two variables: $V_m$ and $\xi_m$
	    (in addition to constant parameters). 
	   In the Appendix we show that the dependence for a coreless moon is given by 
           \be
	   E_{{\rm grav},m}(V_m,\xi_m)= gR^2\rho_w\left({V_m\over g}\xi_m+ \kappa{\xi_m^2\over 2}  \right)  
	   \label{eq:egravsph}
	   \ee
	    (eq. \ref{eq:egravfull}),
	   where $\rho_w$ is the density of water,  $g$ is the surface gravity and
	   \be
	   \kappa\equiv  {2/ 5} \ .
	   \ee
	   Equation (\ref{eq:egravsph}) applies for each $m$, and we do not use the summation convention in this paper. 
	   Minimizing $E_{{\rm grav},m}$ with respect to $\xi_m$ then gives the desired static tidal response:
	   \be
	    \bar{\xi}_m = {-\bar{V}_m/g\over \kappa}   \label{eq:tr}
	   \ee
	   after replacing $V_m\rightarrow \bar{V}_m$ and $\xi_m\rightarrow \bar{\xi}_m$. 

	  One may rephrase the result in terms of Love numbers.  Although we shall not
	  make much  use of Love numbers,  we 
	  mention them to connect to the literature. The $h$-Love number is the ratio
	   $\xi_m/\left(-V_m/g  \right)$, and so the $h$-Love number associated with the
	  static tide in the two-layer model is
	  \be
           h^{(\rm 2-layer)} = {1\over \kappa} = {5\over 2} \label{eq:love0}
	  \ee
	  And the $k$-Love number is the ratio of the self-potential at the surface (eq. \ref{eq:self-potential})
	   to $V_m$, which gives
	   \be
	    k^{(\rm 2-layer)} =  {3\over 2}
	   \ee

	   
	   \subsection{Epicyclic Tide}
	   \label{sec:etide}
	   When we include the epicyclic tide, the total potential is $\bar{V}_m+V_m'$
	    (eq. \ref{eq:vdecomp}), 
	   and we write the response as
	   $\bar{\xi}_m+\xi_m'$. 
	    We  calculated $\bar{\xi}_m$ above by minimizing the gravitational energy.
	    Here we calculate $\xi_m'$ by minimizing the sum of the gravitational and elastic
	    energies.
	    As is apparent from Figure \ref{fig:cartoon}, $\xi_m'$ will depend on  $\gamma_S$ and
	     $\gamma_{\rm ice}$. 
	     In the present subsection,  those two angles are assumed  to be given.
	   
	     The gravitational energy associated with the epicyclic tide, which we label $E_{{\rm grav}}'$, is the difference between the total 
	     tidal energy, and the energy associated with the static tide, i.e., from equation (\ref{eq:egravsph}),
	     \be
	     E_{{\rm grav},m}' &=& 
	    E_{{\rm grav},m}(\bar{V}_m+V_m',\bar{\xi}_m+\xi_m') - E_{{\rm grav},m}(\bar{V}_m,\bar{\xi}_m) 
	    \qquad \\
	     &=& gR^2\rho_w \left(  (V_m'/ g)\xi_m'+\kappa {\xi_m'^2\over 2}
	     +\xcancel{(V_m'/ g)\bar{\xi}_m}
	      \right) \label{eq:e2}
	     \ee
	     where we have
	     used the static tide solution to eliminate two terms linear in primed variables, and
	      ``scratched out'' the term independent 
	    of $\xi_m'$, because it will not play a role in the energy minimization.

	    We turn now to the elastic energy. 
  A thin solid shell that is deformed from an initially spherical state, with radial displacement $\xi_mY_{\ell,m}$, 
  has elastic deformation energy 
	   \be
	   E_{{\rm elas},m} = \mu d \cdot {\xi_m^2} \ , \label{eq:eelas}
	   \ee
	   where $d$ is the shell's thickness and $\mu$ is the rigidity\footnote{
	    We assume  that the Poisson ratio is $\nu=1/3$.
	    For a general $\nu$, one should  replace $\mu\rightarrow \mu 4{1+\nu\over 5+\nu}$ throughout.
	    \label{foot:nu}
	    }.
	    Equation (\ref{eq:eelas}) is derived in
	      \S \ref{sec:elasticenergy} in the Appendix, which is based on  \cite{1947TrAGU..28....1M}.
             To determine the elastic energy associated with the epicyclic tide, which we will label $E_{\rm elas}'$, 
             one must account for the fact that the static tidal shape is not spherical (Fig. \ref{fig:cartoon}a). 
             A naive expectation might  be that the elastic energy stored by stretching the shell from $\bar{\xi}_m$
             to $\bar{\xi}_m+\xi_m'$ would be $E_{\rm elas,m}'=\mu d\cdot \xi_m'^2$. But that  expectation is only 
              correct for the $m=0$ and $m=+$ components. For the $m=-$ component, which 
              quantifies the orientation of the  bulge, 
              one must 
              account for the rotation of the shell. For example, a pure rotation of the
              ice shell without a change in its shape (as depicted in Figs. \ref{fig:cartoon}a$\rightarrow$b) does not produce
              elastic energy, even though it produces a radial displacement $\xi_m'$. 
              Instead, what is relevant is the displacement  relative to the rotated shell \citep{2010Icar..209..631G}.
              In order to quantify that, we 
              first note that if the shell does not rotate, then
              \be
                E_{{\rm elas},-}'=\mu d \cdot (\xi_-')^2 \ , \ 
               {\ \rm for\ } \gamma_{\rm ice}=0 \ . \label{eq:gice0}
               \ee
                In order to adapt that
              to $\gamma_{\rm ice}\ne 0$, 
              we introduce the angle $\gamma_{\rm bul}$, which is defined via
	     \be
	     \xi'_-=2\gamma_{\rm bul}\bar{\xi}_+ \ ; \label{eq:ang}
	     \ee
	      $\gamma_{\rm bul}$ is the angle by which the bulge is tilted away 
	     from the guiding center (Fig. \ref{fig:cartoon}). 
	     As a verification of that assertion, we form the sum of the $m=\pm $ terms 
	    \be  
	  \xi_+Y_{\ell,+}+\xi_-Y_{\ell,-}  &\propto&
	    (\bar{\xi}_++\xi_+')\cos2\phi +  \xi_-'\sin2\phi  \nonumber
	    \\
	    &&\approx   \bar{\xi}_+\cos(2(\phi-\gamma_{\rm bul})) + \xi_+'\cos 2\phi
	    \nonumber
	    \ee
	    where we have used that $\bar{\xi}_-=0$, and  assumed that $|\gamma_{\rm bul}|\ll 1$.
	    One sees that the $\xi_-'$ term reorients the static bulge by the angle $\gamma_{\rm bul}$, as claimed. 
	     Since $\gamma_{\rm bul}$ has a clearer physical interpretation than $\xi_-'$, we use equation (\ref{eq:ang}) to replace the latter
	     with the former. 
	     Equation  (\ref{eq:gice0}) thus becomes
	      	      $E_{\rm elas,-}'=\mu d \cdot (2\gamma_{\rm bul}\bar{\xi}_+)^2$, for $\gamma_{\rm ice}=0$.  In order to generalize
	      that to arbitrary $\gamma_{\rm ice}$, one must simply replace $\gamma_{\rm bul}\rightarrow \gamma_{\rm bul}-\gamma_{\rm ice}$. 
	    Consequently, the elastic energy associated with the epicyclic tide is
	    \be
	    E_{\rm elas,m}'=
	       \mu d \times   \begin{cases} \xi_m'^2 & \text{for } m=0, + \\
	         4(\gamma_{\rm bul}-\gamma_{\rm ice})^2(\bar{\xi}_+)^2 & \text{for } m=-
	        \end{cases}  \qquad
	        \label{eq:eprimeelas}
	        \ee
	    We may now minimize the total epicyclic   energy as a function of the three $\xi_m'$, for $m=0,\pm$.  But before doing so
	    we replace   $\xi_-'$    with $\gamma_{\rm bul}$  in the gravitational energy as well, after which we will be able to minimize the energy as
	    a function of $\xi_0$, $\xi_+$, and $\gamma_{\rm bul}$. 
	    We therefore  separate out the $m=-$ component of $E_{\rm grav}'$ 
	    in equation (\ref{eq:e2}). 
	    For $V_-'$, we observe from the $V$'s  given in equations (\ref{eq:vbar})--(\ref{eq:vprime}) that
	     \be
	     V_-' = 2\gamma_S\bar{V}_+ \label{eq:gs3}
	     \ee
	     where 
	      $\gamma_S$ is Saturn's angular position.
	     Equation (\ref{eq:gs3}) is analogous to
           equation     (\ref{eq:ang}). 
	   The $m=-$ gravitational energy may now be written as
	    \be
	    E_{\rm grav,-}' &=& 2gR^2\rho_w\left({\bar{V}_+/ g}  \right)^2{1\over\kappa}\left(
	    {\gamma_{\rm bul}^2}  -2\gamma_S\gamma_{\rm bul} \right)  \ ,
	    \label{eq:egravmid}
	    	    \ee
		after using again the static response (eq. \ref{eq:tr}).
       One would expect  on physical grounds that
       $E_{\rm grav,-}'  \propto (\gamma_{\rm bul}-\gamma_S)^2$ in the two-layer model, 
       because it should depend only on the misalignment between the bulge and Saturn. 
       And indeed this is consistent with equation (\ref{eq:egravmid}), in view of the fact
       that we may drop terms independent of $\xi_m'$ (or $\gamma_{\rm bul}$). 
     Collecting results, the gravitational energy has now been expressed in terms of our desired variables as
	    	    \be
	   E_{{\rm grav},m}' =  gR^2\rho_w
	   \times  
	   \hspace{4.5cm}
	   \nonumber
	   \\
	    \begin{cases}  (V_m'/ g)\xi_m' +\kappa {\xi_m'^2\over 2}&\quad \text{for } m=0, + \\
 2\left({\bar{V}_+/ g}  \right)^2{1\over\kappa}{\left(
	    \gamma_{\rm bul}^2-2\gamma_{\rm bul}\gamma_S \right)}
	         &\quad \text{for } m=-
	        \end{cases} \qquad
	        \label{eq:eprimegrav}
	    \ee

          We obtain the epicyclic response by 
	   minimizing $E_{\rm grav}'+E_{\rm elas}'$ with respect to $\xi_0',\xi_+'$, and $\gamma_{\rm bul}$, which 
	   results in
		   \begin{align}
	   \xi_m' &= {-V_m'/g\over \kappa+\R}    &\quad {\rm for}\   m=0 ,  + \label{eq:ximprime2} \\
	   \gamma_{\rm bul} &= {\kappa \gamma_S+\R \gamma_{\rm ice} \over \kappa+\R}     &{\rm for}\   m=-
	   \label{eq:ximprime3}
	   \end{align}
	   where
	   \be
	   {\R}\equiv {2\mu d\over g R^2\rho_w} \label{eq:hardness} \ .
	   \ee
	   We call  the dimensionless  $\R$ the   
	      ``hardness parameter.''  It plays an  important role.
	    For a hard shell, ${\cal R}\gg 1$, and equation (\ref{eq:ximprime3}) implies that $\gamma_{\rm bul}\approx \gamma_{\rm ice}$
	    (Fig. \ref{fig:cartoon}b). And for a shoft shell, ${\cal R}\ll 1$, and $\gamma_{\rm bul}\approx \gamma_S$ (Fig. \ref{fig:cartoon}c).
	     \cite{2010Icar..209..631G} introduced ${\cal R}$, albeit  with slightly different
	    order-unity constants. 
	     It is analogous to
	   the 
	    ``effective rigidity'' of a core 	   \citep{love,1999ssd..book.....M}, but for the shell. 
	    The hardness parameter  is comparable
	   to the ratio of elastic energy available in the shell $(\sim \mu d R^2)$ to the gravitational 
	   energy of the moon $(\sim \rho_w R^3 g R)$. 
	   Thin shells are soft, and thick shells are hard, with the transitional thickness being
	   \be
	   d_{\rm hsb}&\equiv& {gR^2\rho_w\over 2\mu}  \\
	   &=& 0.83 {\rm km} \ ,
	   \label{eq:dhsb}
	   \ee
	   where subscript hsb stands for hard-soft boundary, and the numerical expression is for Enceladus.
	   The value of $d_{\rm hsb}$ scales with the mass of the body; for Europa, 
	   $d_{\rm hsb}=$ 380km, and for
	   Titan $d_{\rm hsb}=$ 1100km. 
	   We also see from equation (\ref{eq:ximprime2}) that the $h$-Love number    associated with 
	    the epicyclic tide is $1/(\kappa+\R)$, for $m=0,+$. That means that the body acts as a fluid
	   for $\R\ll 1$, and the deformation is reduced relative to a fluid for $\R\gg 1$.

	   With $\gamma_{\rm bul}$ in hand, we may now determine the torque.
	    The torque on the bulge is
	   \be
	   T&=&-{\dd E_{\rm grav,-}'\over \dd \gamma_{\rm bul}}  \\
	   &=& 4gR^2\rho_w \left( {\bar{V}_+/ g} \right)^2{1\over \kappa}(\gamma_S-\gamma_{\rm bul})
\\
	   &=& 4gR^2\rho_w \left( {\bar{V}_+/ g} \right)^2{\R\over \kappa(\kappa+\R)}(\gamma_S-\gamma_{\rm ice})   \label{eq:torquenocore}
	   \ee
	   	   An alternative way to calculate the torque is to integrate the specific torque from Saturn
	   ($\dd_\phi V$), multiplied by the density perturbation at the surface (eq. \ref{eq:rhopapp}). 
	   That  leads to the same expression for $T$. 
	   To order of magnitude, the torque can be understood 
	    in the hard shell limit ($\R\gg 1$) as follows: the specific tidal force is $\bar{V}_+/R$; the lever arm is $R(\gamma_S-\gamma_{\rm ice})$; and the mass in the bulge is $\bar{\xi}_+/R$ times the mass of the moon ($\rho_wR^3$). 
	   Multiplying those four quantities gives $T\sim gR^2 \rho_w (\bar{V}_+/g)^2(\gamma_S-\gamma_{\rm })$, 
	   after inserting the height of the static bulge from equation (\ref{eq:tr}). 
	   For soft shells the torque is reduced, because in 
	   that case the bulge nearly points at Saturn.
	   It  is reasonable that the torque only depends on the relative angle $\gamma_S-\gamma_{\rm ice}$, 
	   whether the shell is hard or soft, because
	   of the overall rotational invariance of the setup.

	   We will use the
	    torque to write the equation of motion for the shell, and then solve for $\gamma_{\rm ice}(t)$. 
	    But before doing so, we shall turn to the three-layer model.
	   One may note that we neglect two other torques: the pressure torque from the ocean, 
	   and the elastic torque from the shell. Those  are internal torques, and so cancel each 
	   other.  The cancellation is a consequence of the implicit assumption that 
	   the  ocean is able to react sufficiently quickly that the lowest energy state is reached at given values of $\gamma_S$ and $\gamma_{\rm ice}$.
	  An additional subtlety is that the torque from Saturn acts directly on the ocean, rather than on the shell.
	  But the pressure torque from the ocean on the shell is equal to the torque from Saturn 
	  \citep{2010Icar..209..631G}.

	 \section{Three-Layer Model}
	 \label{sec:threelayer}
\begin{figure}[h]
\centering
    \includegraphics[width=.48\textwidth]{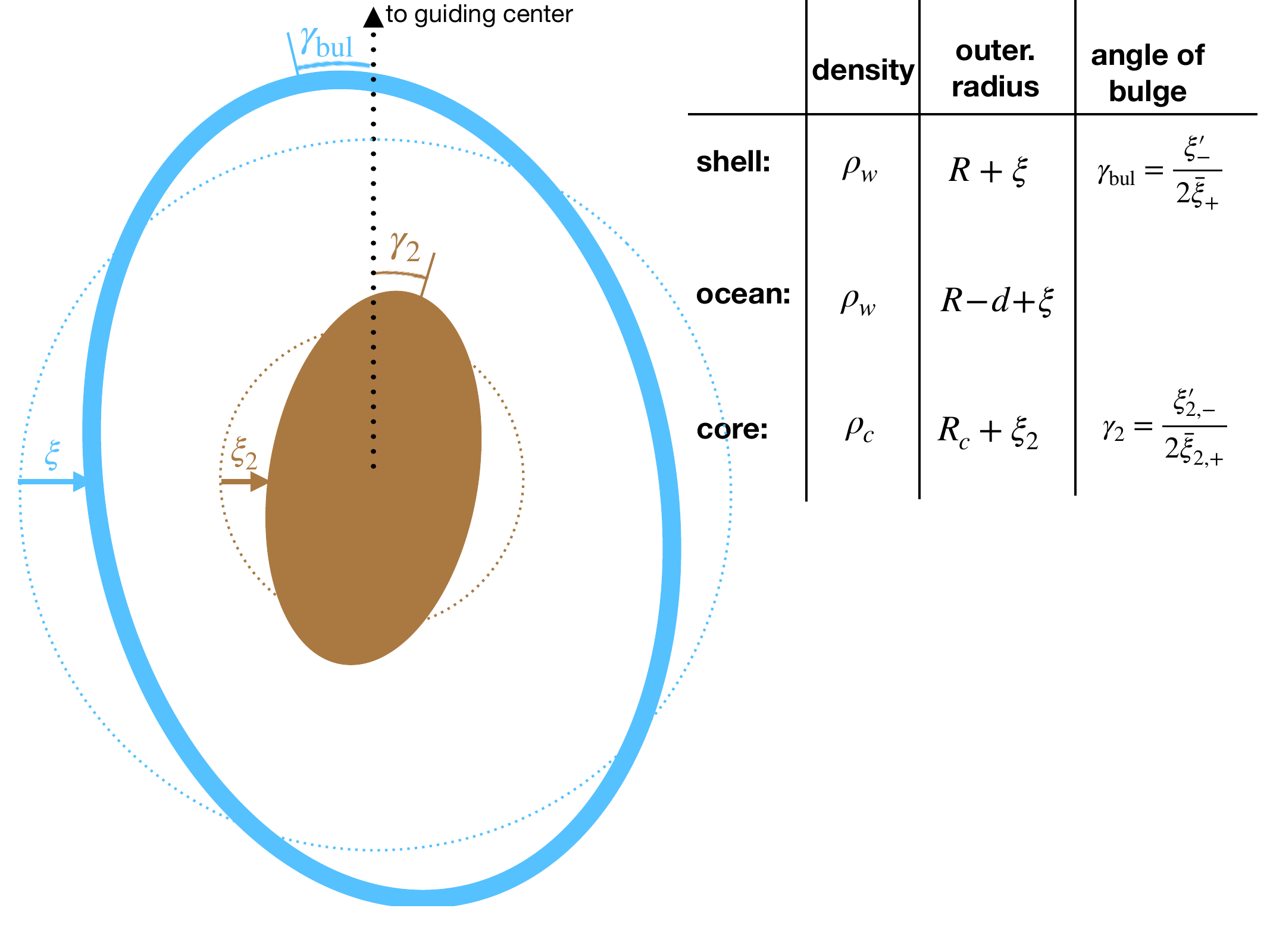}
\caption{{\bf Three-layer model:}  The dotted circles represent the unperturbed (i.e., spherical) Enceladus. 
}
\label{fig:core}
\end{figure}

	 We now include the core. The three-layer model is sketched in Figure \ref{fig:core}.
	The radial displacement of the shell relative to a spherical shape is decomposed as before (eq. \ref{eq:xiexp}), 
	and that of the core is\footnote{
	The subscript  2 always refers to the core (not to $m=+2$), 
	 and we will use a subscript 1 to refer to the ice shell when needed. 
	   }
	 	   \be
	   \xi_2&=& \xi_{2,0}Y_{\ell,0}+\xi_{2,+}Y_{\ell,+}+\xi_{2,-}Y_{\ell,-} \label{eq:xi2}
	   \ee
	 Our basic assumptions for the core are that  its shape is determined by the static tide, for
	 which elasticity is assumed to play no role, i.e., it behaves as a fluid in reaction to the static
	 tide. 
	 And for the epicyclic tide, we  allow it to rotate as a solid body.
          Our development here closely parallels the two layer model in \S \ref{sec:toy}, and so
          we skip  some of the explanations that may be found in the previous section.
          The reader uninterested in the order-unity corrections introduced by the three-layer
          model may wish to skip  to \S \ref{subsec:gamice}. 
	 
	   \subsection{Static Tide}
	   \label{sec:threestat}
	   
	   From equation (\ref{eq:egravfull}), the gravitational energy is
	   \begin{equation*}
	   \begin{aligned}
	   E_{{\rm grav},m}(V_m,\xi_{m},\xi_{2,m}) = gR^2\rho_w \times \hspace{4cm}  \\
	    \left( {V_m\over  g}\xi_{m}+ \kappa_{2V}{V_m\over  g}\xi_{2,m}
	   +\kappa_{11}{\xi_{m}^2\over 2}+\kappa_{22}{\xi_{2,m}^2\over 2}+\kappa_{12}\xi_{m}\xi_{2,m}
	    \right) 
	    \end{aligned}
	   \end{equation*}
	   where the dimensionless $\kappa$ coefficients are order-unity, and are defined  below equation (\ref{eq:egravfull})
	   in the Appendix in terms of parameters shown in Figure \ref{fig:core}. 
	   The numerical values of these coefficients for Enceladus are listed in Appendix
	   \ref{sec:numbers}.

	   Minimizing $E_{{\rm grav},m}$ with respect to $\xi_{m}$ and $\xi_{2,m}$
	   yields  the linear set of equations
	   \be
	  \left( \begin{array}{cc}
	   \kappa_{11}&\kappa_{12} \\
	   \kappa_{12}&\kappa_{22}
	   \end{array}\right)
	    \left( \begin{array}{c}
	    \bar{\xi}_{m}/(-\bar{V}_m/g) \\
	    \bar{\xi}_{2,m}/(-\bar{V}_m/g)
	   \end{array}\right) = 
	     \left( \begin{array}{c}
	       1 \\ \kappa_{2V}
	   \end{array}\right) 
	   \label{eq:heq}
	   \ee
	   which is the generalization of equation (\ref{eq:tr}). 
	   The solution for the $\bar{\xi}$'s is obtained by inverting the matrix in equation (\ref{eq:heq}).
	   	   The inversion is trivial, but a little messy to express explicitly.  In what follows, we assume
	   the inversion has been done, and parameterize the result via the  numbers $h$ and
	   $h_2$, 
	   \be
	   	   \left( \begin{array}{c}
	    h \\
	    h_2
	   \end{array}\right) 
	   \equiv 
	   \left( \begin{array}{c}
	    \bar{\xi}_{m}/(-\bar{V}_m/g) \\
	    \bar{\xi}_{2,m}/(-\bar{V}_m/g)
	   \end{array}\right) 
	    \label{eq:heq2}
	   \ee
	   Numerical values for the $h$'s are also  in 
	   Appendix \ref{sec:numbers}.
	   We note that $h$ is the $h$-Love number for   the static tide.
	   And the $k$-Love number is, 
	   using equation (\ref{eq:self-potential}) and the corresponding internal
	   potential produced by the core,
	   \be
	    k &=& h-1 \ .
	   \ee 
	   This relationship is  true for any fluid body, 
	   as it is equivalent to the surface being an equipotential.
	   With our numerical values, we get $k=0.92$ for Enceladus, 
	   which is in reasonable agreement with 
	   \cite{2016Icar..277..311V}, who get  0.989 for their fluid $k$-Love number.
	   We shall not use $k$ further.

           \subsection{Epicyclic Tide}
           \label{sec:threeepi}
	   The generalization of equations   (\ref{eq:eprimeelas}) \& (\ref{eq:eprimegrav}) for the two energies is
	          \be
	       	    E_{\rm elas,m}'&=&
	       \mu d \times   \begin{cases} \xi_{m}'^2 & \text{for } m=0,+ 
	       \\
	         4(\gamma_{\rm bul}-\gamma_{\rm ice})^2h^2\left({\bar{V}_+/ g}  \right)^2 & \text{for } m=-
	        \end{cases} \qquad
\ee
and
\be
	   E_{{\rm grav},m}' &=&
	    E_{{\rm grav},m}(\bar{V}_m+V_m',\bar{\xi}_{m}+\xi_{m}',\bar{\xi}_{2,m}+\xi_{2,m}') - 
	    \nonumber
	    \qquad
	    \\ &&\qquad E_{{\rm grav},m}(\bar{V}_m,\bar{\xi}_{m},\bar{\xi}_{2,m}) 
\ee
\vspace{-.5cm}
\be	    
	 &=&    gR^2\rho_w  \times  \nonumber  \\
	   && \begin{cases}  
	     {V_m'\over  g}\xi_{m}'
	     +\kappa_{11}{\xi_{m}'^2\over 2} + \xcancel{{V_m'\over  g}\left(\bar{\xi}_2+\kappa_{2V}\bar{\xi}_{2,+}\right)}
  & \text{for } m=0,+ \\
  2\left( {\bar{V}_+/ g} \right)^2 K 	         & \text{for } m=-
	        \end{cases}
	  \qquad  
	    \ee
	         where 
	         \be
	        K \equiv
	   \kappa_{11}h^2{\gamma_{\rm bul}^2 } -2h\gamma_S\gamma_{\rm bul}  
  +2\kappa_{12}hh_2 \gamma_{\rm bul}\gamma_2
	     + 
	     \nonumber \\
	     \kappa_{22}h_2^2{\gamma_2^2}	-2\kappa_{2V}h_2\gamma_S\gamma_2   \ ;   
	      \ee
	         the  bulge angles  $\gamma_{\rm bul}$
	          and $\gamma_2$ are defined in Figure \ref{fig:core}, 
         in accordance with equation (\ref{eq:ang});  we have set $\xi_{2,m}'=0$ 
         for $m=0$ and $m=+$ because those two components are unaffected
         by the assumed solid-body rotation of the core; and we have used the relations
         for the static tide.

          We solve for  the epicyclic tidal response by 
          minimizing   $E_{\rm grav,m}'+E_{\rm elas,m}'$ with respect to $\xi_{m}'$ for $m=0, +$, and with respect
          to $\gamma_{\rm bul}$ for $m=-$,  yielding
	   \be
	   \xi_{m}' &=&  {-V_m'/g\over \kappa_{11}+\R}
	  \ \ \  {\rm for\ } m=0, +  \label{eq:ximprime}
	   \\
	   \gamma_{\rm bul} &=& {1/h\over \kappa_{11}+\R}\left(
	   \R h \gamma_{\rm ice}-h_2\kappa_{12}\gamma_2
	     +\gamma_S
	    \right)
		\label{eq:gam1sol}
	   \ee
	   Comparing with the two-layer version of these equations (eqs. \ref{eq:ximprime2}--\ref{eq:ximprime3}), 
	   we see that the only changes are to the order-unity coefficients, as well as the contribution of
	    $\gamma_2$ to $\gamma_{\rm bul}$, which comes about because the shell now feels the potential
	    from a rotated tidal bulge.

          \subsection{Solution for $\gamma_{\rm ice}(t)$}
          \label{subsec:gamice}
         Our next step is to calculate the torques on the shell and core, and use those to solve the
         equations of motion for $\gamma_{\rm ice}(t)$ and $\gamma_2(t)$.  We do that in Appendix \ref{sec:corelib}. 
         Here, in order to reduce algebraic complexity,  we anticipate that the rotation of the core
         may be neglected for Enceladus, and so we set $\gamma_2\rightarrow 0$. 
         We will verify shortly that dropping $\gamma_2$ does not affect the result for Enceladus. 
          Setting $\gamma_2\rightarrow 0$ in the energies, the torque on the shell is
         \be
         T &=& -{\dd E'_{\rm grav,-} \over \dd \gamma_{\rm bul}} \\
                  &=& \hat{T}\left({\gamma_S\over h\kappa_{11}}-\gamma_{\rm ice}  \right)
         \ee
         where
         \be
         \hat{T}&\equiv &4 gR^2\rho_w(\bar{V}_+/g)^2{h^2\kappa_{11}\R\over \kappa_{11}+\R} 
         \label{eq:hatt0}
         \\
         &=&
         n^2{n^2R\over g}  {12\pi \over 5}{\rho_w}R^5{h^2\kappa_{11}\R\over \kappa_{11}+\R}
         \label{eq:hatt}
         \ee
         after using equation (\ref{eq:gam1sol}) to replace   $\gamma_{\rm bul}$ with $\gamma_{\rm ice}$, 
         and the second equality follows after substituting in for  $\bar{V}_+$  from equation (\ref{eq:vbar}).
         Comparing with the torque in the two layer model (eq. \ref{eq:torquenocore}), we see that the main effect of the core is to reduce
         Saturn's effective angular position from $\gamma_S\rightarrow\gamma_S/(h\kappa_{11})$. 
         This is sensible because the shell now feels the quadrupolar potential from both the core and
         Saturn, and so with the core's bulge pointing at the guiding center, the net effect is as
         though
          Saturn's angular displacement was decreased.

\begin{figure}[h]
\centering
    \includegraphics[width=.45\textwidth]{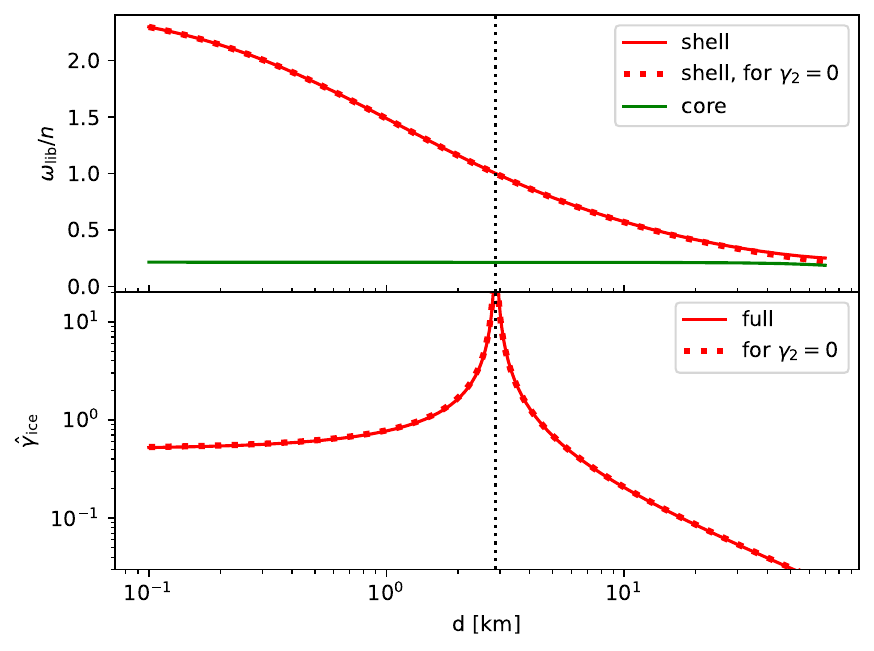}
\caption{{\bf Libration Frequencies and Amplitude of Forced Libration:}  
Solid curves include libration of the core, and dotted curves neglect it.
The dotted vertical line is at the resonant thickness of the shell.
}
\label{fig:lib}
\end{figure}

	   The equation of motion of the shell is\footnote{
	   To be more accurate, one should also include the time-dependence of the moment of inertia
	   \citep{2013Icar..226..299V}. We find that that effect produces corrections that are
	   of order $\epsilon\equiv g R\rho_w/\mu$, which is $\ll 1$ for Enceladus, but is
	   of order unity for Titan and Europa  (see footnote \ref{foot:eps}).
	   }
	   \be
	    C_{\rm ice}\ddot{\gamma}_{\rm ice} &=& T 
	     \\
                  &=& \hat{T}\left({\gamma_S\over h\kappa_{11}}-\gamma_{\rm ice}  \right)
	   \ee
	   where 
	      \be
           C_{\rm ice}= {8\pi\over 3}\rho_wR^4d
           \ee
           is the moment of inertia of a thin spherical shell. 
           Thus,
           \be
         \ddot{\gamma}_{\rm ice} + \omega_{\rm lib}^2 \gamma_{\rm ice} = \omega_{\rm lib}^2 
         {2e\over h\kappa_{11}}  \sin n t
         \label{eq:ddot}
           \ee
         where the frequency of free librations is given by
         \be
        { \omega_{\rm lib}^2\over n^2} &=&{\hat{T}\over C_{\rm ice}}
        \\ &=&  {9 \over 10} {n^2R\over g} {R\over d}{h^2\kappa_{11}\R\over \kappa_{11}+\R} \label{eq:omlib}  \ .
         \ee
         We note that the core's effect enters only via
            parameters $h$ and
         $\kappa_{11}$. 
         The forced solution to  equation (\ref{eq:ddot}) is 
          \be
          \gamma_{\rm ice} = {\omega_{\rm lib}^2\over \omega_{\rm lib}^2-n^2}{2e\over h\kappa_{11}} \sin n t \ .
          \label{eq:gamice}
          \ee
        Figure \ref{fig:lib} shows the libration frequency from equation (\ref{eq:omlib}), versus $d$, as the dotted
        curve in the top panel.  And it shows the libration amplitude from equation (\ref{eq:gamice}) as the dotted
        curve in the lower panel.  Also shown in the top panel as solid curves are the frequencies of the two free libration
        modes, as determined by solving the  coupled equations of motion for both shell and core in Appendix 
        \ref{sec:corelib}.
        It may be seen that including the libration of the core has negligible effect, justifying
        our neglect of it.   
        Comparing this figure with the corresponding one in 
        \cite{2016Icar..277..311V} (their Fig. 2) shows that the agreement
        is better than 20\%. 
        
        More significantly, the figure shows that 
        Enceladus's shell experiences a libration resonance at a  thickness
           \be
         d_{\rm res} = 2.8 {\rm km}  \ , 
                  \ee
         which follows from setting $\omega_{\rm lib}/n$ to unity for Enceladus's parameters.
        The libration resonance is one of the main drivers of Enceladus's limit cycle. 
        At  libration resonance, the forced libration amplitude appears to  diverge. But the
        divergence is unphysical. In the next section, we will remove it  by 
        modelling more carefully what happens near resonance.

\section{Heating Rate}    
\label{sec:heatingrate}                
We turn finally to the tidal heating rate. Our basic assumption is that the heating rate is 
proportional to the time-averaged elastic energy  in the epicyclic tide, i.e., 
\be
H
 = {2n\over Q_{\rm ice}}\langle E_{\rm elas}' \rangle \ , \label{eq:heat0}
\ee
where the quality factor $Q_{\rm ice}$ is a constant.  
We continue to assume here that the core does not librate, based on the agreement in Figure \ref{fig:lib}. 
We sum the three $m$ components of $E_{{\rm elas},m}'$, 
and insert the tidal response (eqs. \ref{eq:ximprime}--\ref{eq:gam1sol}), along with 
the expressions for the potential (eqs. \ref{eq:vbar}--\ref{eq:vprime}), which yields
\be
E_{\rm elas}' = {6\pi\over 5}{n^4R^6\rho_w\over g}{\R\over (\kappa_{11}+\R)^2} \times 
\qquad\qquad\qquad\qquad
\nonumber
\ee
\vspace{-.4cm}
\be
\left(
3 e^2\cos^2 n t + \left( \gamma_S-h\kappa_{11}\gamma_{\rm ice} \right)^2
\right) \label{eq:eelasplast}
\ee
The first term in the big brackets is from the radial tide, and the second is from the librational tide. 
If $\gamma_{\rm ice}$ is neglected, the ratio of the time-averaged energies in those two 
tides is 3:4 \citep{1999ssd..book.....M}.
We next insert for $\gamma_{\rm ice}$ from equation (\ref{eq:gamice}), and take the time average;
 equation (\ref{eq:heat0}) then implies
\be
H
={2n\over Q_{\rm ice}} {21\pi \over 5}{n^4R^6\rho_w\over g}{\R\over (\kappa_{11}+\R)^2}e^2 \times 
\hspace{2cm} \nonumber \\
\left({3\over 7}+{4\over 7}\left( {n^2 \over \omega_{\rm lib}^2-n^2} \right)^2  \right) \ , \hspace{.5cm}
\label{eq:heat}
\ee
where the 3/7 term is from the radial tide and the 4/7 term from the librational. 

     This is nearly the final heating rate. But 
	     the divergence of the librational tide at resonance ($\omega_{\rm lib}= n$) is not physical. 
	   It must be fixed in order to model the shell's evolution. 
	     We  account for two corrections that prevent divergence at resonance, dissipation and nonlinearity. 
	     Both corrections play a role in the limit cycle in Goldreich et al. Dissipation is the
	     dominant correction when the shell gradually grows towards its resonant thickness during 
	     the resonant libration stage; and nonlinearity dominates when the shell swiftly melts
	     during the runaway melting stage.

	    Dissipation provides a finite linewidth, which we model
	       by adding a damping term  to the equation of motion (eq. \ref{eq:ddot}):
            \be
	    \ddot{\gamma}_{\rm ice}+2\eta\dot{\gamma}_{\rm ice}+ \omega_{\rm lib}^2 \gamma_{\rm ice} = \omega_{\rm lib}^2
	    {2e\over h\kappa_{11}} \sin nt \hspace{1cm}
	     \label{eq:d2g}
	    \ee
	    	    where the damping factor $\eta$ is to be determined.
	   The  forced amplitude  is now changed to
	   \be
	            \hat{\gamma}_{\rm ice} = {\omega_{\rm lib}^2\over  \sqrt{(\omega_{\rm lib}^2-n^2)^2 + (2\eta n)^2}}{2e\over h\kappa_{11}} \sin n t  \nonumber
	             \ee
	             \newline
          which is 	          
	    the same as previously, 
	   except with the replacement
	   \be
	   \omega_{\rm lib}^2-n^2 \rightarrow \sqrt{(\omega_{\rm lib}^2-n^2)^2 + (2\eta n)^2} \label{eq:sub}
	   \ee
           in the denominator. 
	   	   After  inserting this new forced  $\gamma_{\rm ice}(t)$ into equation
	   (\ref{eq:eelasplast}) and taking the time average, the heating rate 
	   is the same as 
	   before (eq. \ref{eq:heat}), aside from the same replacement (eq. \ref{eq:sub})   in the denominator of the librational tide term (in addition to a small $O(\eta^2/n^2)$ correction to the numerator
	   that we drop). 
	   
	           To evaluate $\eta$, we equate the energy dissipation rate that results from equation 
        (\ref{eq:d2g}) to the rate $H$ in  equation (\ref{eq:heat}) near resonance,  after 
        making the substitution  of equation (\ref{eq:sub}). 
        The former rate is 
         $\eta 2C_{\rm ice}\langle \dot{\gamma}_{\rm ice}^2 \rangle=\eta C_{\rm ice}n^2 \hat{\gamma}_{\rm ice}^2$, 
         which thereby yields
       \be
       \eta = {1\over C_{\rm ice}n^2\hat{\gamma}_{\rm ice}^2}H 
	    &\approx & 
		  {1\over 2} {n\over Q_{\rm ice}}  
            {\kappa_{11} \over \kappa_{11}+\R}	  \label{eq:eta} \ .
       \ee
       where the approximate form is for $\omega_{\rm lib}\approx n$. 
       That completes the correction to $H$ due to dissipation.

       The second correction to equation (\ref{eq:heat}) is due to nonlinearity. We have assumed that $\hat{\gamma}_{\rm ice}\ll 1$, and so
       if $\hat{\gamma}_{\rm ice}\sim 1$ we will have overestimated the heating rate.
      We account for nonlinearity
      in a rough way by using an effective damping factor, $\eta_{\rm nonlin}$. 
      We start from the damped equation of motion (eq. \ref{eq:d2g}), with $\eta\rightarrow \eta_{\rm nonlin}$. 
      We then determine $\eta_{\rm nonlin}$ by requiring that $\hat{\gamma}_{\rm ice}=1$ at exact resonance
      ($\omega_{\rm lib}=n$), which implies that
      \be
      \eta_{\rm nonlin}=n  {e\over h\kappa_{11}}  \label{eq:etanonlin}
      \ee
      Finally, we modify the replacement rule (eq. \ref{eq:sub}) to add both $\eta$'s in quadrature. 
   
      Collecting results, the final heating rate is
      \be
H={2n\over Q_{\rm ice}} {21\pi \over 5}{n^4R^6\rho_w\over g}{\R\over (\kappa_{11}+\R)^2}e^2 \times
\nonumber \qquad\qquad\qquad \\
\left({3\over 7}+{4\over 7} {n^4\over
({\omega_{\rm lib}^2}-n^2)^2 + (2\eta n)^2+(2\eta_{\rm nonlin}n)^2  }  \right)  \qquad
\label{eq:heatfin}
\ee
\ \newline
\noindent
with $\R$ from  eq. (\ref{eq:hardness}), $\omega_{\rm lib}$ from eq. (\ref{eq:omlib}), $\eta$ from eq. (\ref{eq:eta}), and $\eta_{\rm nonlin}$ from eq. 
(\ref{eq:etanonlin}).
         
       The expression  simplifies in the hard shell limit ($\R\gg 1$), 
       which for Enceladus is applicable when the shell is much thicker than a  kilometer (eq. \ref{eq:dhsb}). 
       In that case,
       \be
       H\approx 
       {21\pi \over 5}{\rho_w^2R^8n^5\over \mu Q_{\rm ice}}{e^2\over d} \times \nonumber 
       \qquad\qquad\qquad\qquad\qquad\qquad \\
\left({3\over 7}+{4\over 7} {1 \over
\left({\omega_{\rm lib}^2\over n^2}-1\right)^2 +\left({1\over Q_{\rm ice}}{\kappa_{11}\over \R}\right)^2+ \left({2e \over h\kappa_{11}}\right)^2  }  \right)  \qquad
\label{eq:heatsimp}
       \ee
       \newline
       for $\R\gg 1$.
       If the shell  also has $\omega_{\rm lib}\ll n$, which for Enceladus occurs when the shell is 
      much thicker than $d_{\rm res}=2.8$ km, the main bracketed term in the above expression is nearly equal to unity.

       The heating rates from equations (\ref{eq:heatfin}) and (\ref{eq:heatsimp}) are the main 
       result of this paper. They are plotted in Figure 2 of
       Goldreich et al, where they are seen to agree with each other for $d\gtrsim 1.5$km.
       
\section{Summary}       
       
In this paper, we have derived simple analytic expressions for the following quantities:
\bi
	\item Enceladus's response to the static tide (eq. \ref{eq:heq}).
	\item The forced librations of its shell (eq. \ref{eq:gamice}).
	\item  The tidal heating rate (eq. \ref{eq:heatfin}, which simplifies to 
	eq. \ref{eq:heatsimp} for hard shells).
\ei
These quantities are used in the model Goldreich et al, in which Enceladus is found to experience limit cycles.  We refer the reader to that
paper to more extensive discussions of the implications. 

Although we have aimed for $\lesssim 20\%$ inaccuracies within the context of our model, 
there are a variety of systematic uncertainties.  For example, we assumed that the unperturbed
shell thickness is spherically symmetric, whereas it is known that the ice is thinner
near the south pole region. 
But likely our most suspect assumption is that dissipation in the ice is characterized by a 
quality factor ($Q_{\rm ice}$), of unknown magnitude.  Most of the dissipation is expected to 
happen at the bottom of the ice shell, where the ice is slushy. Modelling the dissipation in 
the slush  accurately is likely challenging.  
       

\ \newline
We thank Peter Goldreich for extensive discussions. We acknowledge NASA grant 80NSSC23K1262.

\appendix
\section{Energies}
In this appendix, we adopt the three-layer model, sketched in Figure \ref{fig:core}.
We
calculate  the gravitational and elastic energies when (i) Enceladus's shape 
is distorted relative to spherical and (ii) a tidal potential is applied. 
The energies derived here are relative to the energy in the spherical state.

\subsection{Gravitational Energy}

	We apply a single   potential component ($V\rightarrow V_mY_{\ell,m}$)  to 
     Enceladus, where $V$ is the potential at Enceladus's surface eq. \ref{eq:vexp}), 
     which has the three $m$ components $V_0,V_+$, and $V_-$. 
     And we set the displacement of the ice shell to 
     $\xi_{m}Y_{\ell,m}$, and the displacement of the core's surface to 
     $\xi_{2,m}Y_{\ell,m}$ (eqs. \ref{eq:xiexp} \& \ref{eq:xi2}). Note that there is no 
     sum on $m$;  that the $m$ of the displacements are taken to be the same
     as the potential, because different $m$'s do not couple; and that the  subscript 2
     always
     refers to the core. 
     Since we assume that the ocean and core each have constant densities, 
     and that the ice shell is very thin, the energy will be determined 
     by the  coefficients $V_m$, $\xi_{m}$, and $\xi_{2,m}$.
   
   The displacement
    $\xi_{m}$ determines the perturbed density field near the surface ($\rho'$); that density 
    is approximately
    \be
    \rho'\approx \rho_w\xi_{m}Y_{\ell,m}\delta(r-R) 
    \label{eq:rhopapp}
    \ee
    In this appendix primes represent deviation from sphericity,  in contrast to the body 
    of the paper where they represent the deviation from the static tide. 
    This density field  produces three contributions to the energy (ignoring for now 
     the displacement
    of the core's surface, $\xi_2$).
    First, from the attraction of $\rho'$ to Saturn,
          \be
      E_{{\rm grav},m}^{\rm (1)} &=& \int \rho' V d\Omega r^2 dr 
      = R^2\rho_w \xi_{m}  V_m \ ,
      \label{eq:egrav1}
      \ee
      where $d\Omega=\sin\theta d\theta d\phi$.
       Second, 
       the gravitational energy due to the interaction of $\rho'$ with itself is
     \be
     E_{{\rm grav},m}^{(2)} ={1\over 2} \int \rho' V' d\Omega r^2 dr  \ ,
     \ee
     where $V'$ satisfies Poisson's equation:
     \be
     \nabla^2 V' = 4\pi G \rho' \ ,
     \ee
     which has solution 
     \be
     V'=-{4\pi\over 5} G\rho_w R\xi_{m}Y_{\ell,m}
     \label{eq:self-potential}
     \ee
      at $r=R$. 
     We therefore have
     \be
     E_{{\rm grav},m}^{(2)}= -{4\pi\over 5}GR^3\rho_w^2   {\xi_{m}^2\over 2}
     \label{eq:egrav2}
     \ee
      And third, from the attraction of $\rho'$ to Enceladus, the potential energy energy (per unit
      volume) to raise the surface is $\rho_w g\xi$, where $g$ is the surface gravity. Integrating that
      over the volume of the distorted surface gives
      \be
      E_{{\rm grav,m}}^{(3)} = gR^2\rho_w{\xi_{m}^2\over 2}
      \label{eq:egrav3}
      \ee

      Turning now to the nonspherical shape of the core (i.e., $\xi_{2,m}$), 	it contributes
	 three more
	terms to $E_{{\rm grav},m}$ that are nearly the same to those produced by $\xi_{m}$  alone
	(eqs. \ref{eq:egrav1}, \ref{eq:egrav2} \& \ref{eq:egrav3}), except that one must make the following replacements
	\be
	\xi_{m}\rightarrow \xi_{2,m} \ , \ \  \rho_w\rightarrow \rho_{cw}\equiv \rho_c-\rho_w \ , \ \ 
	R\rightarrow R_c \ , \ \  {\rm and} \ \ 
	V_m\rightarrow V_m{R_c^2/R^2}
	\ee
	See Figure  \ref{fig:core} for the definition of the symbols.
	 There is also a seventh term that comes from
	the interaction energy between $\xi$ and $\xi_2$. 
        Adding the seven terms gives
	         \be
	         	   E_{{\rm grav},m} = gR^2\rho_w \left( (V_m/ g)\xi_{m}+ \kappa_{2V}(V_m/ g)\xi_{2,m}
	   +\kappa_{11}{\xi_{m}^2\over 2}+\kappa_{22}{\xi_{2,m}^2\over 2}+\kappa_{12}\xi_{m}\xi_{2,m}
	    \right) \label{eq:egravfull}
          \ee
          where 
          \be
             \kappa_{11}&=&   1  -{3\over 5}{\rho_w\over\bar{\rho}} \label{eq:k11}  \\
             \kappa_{22}&=&  {\rho_{cw}\rho_c\over\bar{\rho}\rho_w}   {R_c^3\over R^3}\left(1
		    -{3 \over 5}{\rho_{cw}\over\rho_c}\right)  \label{eq:k22} \\
	     \kappa_{12}&=&  - {R_c^4\over R^4} \left({3\over 5} {\rho_{cw}\over\bar{\rho}}
		    \right)  \label{eq:k12} \\
		                             \kappa_{2V}&=&  {\rho_{cw}\over\rho_w}{R_c^4\over R^4}  \label{eq:k2V}  \\
		    \rho_{cw}&=& \rho_c-\rho_w \\
		    \bar{\rho}&=&\rho_{cw}{R_c^3\over R^3}+\rho_w 
          \ee
          For the special case of a coreless moon one may set $R_c\rightarrow 0$, whence equation (\ref{eq:egravsph}) follows. 

\subsection{Elastic Energy}
\label{sec:elasticenergy}

We consider a thin solid spherical shell of radius $R$ that is displaced by a radial distance $\xi_mY_{\ell,m}$. 
\cite{1947TrAGU..28....1M} show that, for an $m=0$ displacement, Hooke's law implies that the
stress tensor that results has components
	\be
	  \sigma_{\theta\theta}&=&{\xi_0\over R}\mu{1+\nu\over 5+\nu}  {1\over 2}\sqrt{5\over\pi} \left( 3\cos^2\theta+1 \right) \\
	  \sigma_{\phi\phi}&=& {\xi_0\over R} \mu{1+\nu\over 5+\nu}  {1\over 2}\sqrt{5\over\pi} \left( 9\cos^2\theta-5 \right)
	\ee
	and $\sigma_{\theta\phi}=0=\sigma_{\phi\theta}$, where $\mu$ is the rigidity and 
	$\nu$ is the Poisson ratio, adopting the notation of \cite{2008Icar..195..459M}.
	The resulting non-vanishing components
	of the strain tensor are $u_{\theta\theta}=\left( \sigma_{\theta\theta}-\nu\sigma_{\phi\phi} \right)/(2\mu(1+\nu))$, 
	 $u_{\phi\phi}=\left( \sigma_{\phi\phi}-\nu\sigma_{\theta\theta} \right)/(2\mu(1+\nu))$
	 \citep{Landau:1986aog}.
	 The $m=0$ contribution to the elastic  energy is therefore
	 \be
	 E_{{\rm elas},m=0}
&=& {1\over 2}\int \left(\sigma_{\theta\theta}u_{\theta\theta}+\sigma_{\phi\phi}u_{\phi\phi}  \right) d\Omega r^2dr \\
	  &=& \mu d {4(1+\nu)\over 5+\nu}  ( \xi_0)^2
	 \ee
	 where $d\ll R$ is the shell's thickness. 
      
         For the other two values of $m$, the elastic energy must be the same, after replacing $\xi_0\rightarrow \xi_m$. 
         We thereby arrive at equation (\ref{eq:eelas}),  after setting $\nu=1/3$.

  \section{Including Core Libration}
\label{sec:corelib}
We generalize the calculation in \S \ref{subsec:gamice} to account for the libration of the core.
	  The torques on the  shell and core are  
         \be
                   \begin{pmatrix}
              T \\  T_2
           \end{pmatrix}&=& 
           - \begin{pmatrix}
              \dd/\dd \gamma_{\rm bul} \\  
              \dd/\dd\gamma_2\
           \end{pmatrix}
           E_{\rm grav,-}'\Big\vert_{\gamma_{\rm bul}\rightarrow {\rm eq. }\ref{eq:gam1sol}}
           \\ &=&   {\hat{T}\over h \kappa_{11}}\left(  
            \bld{\tau_V}{\gamma_S}
           - \bld{\tau}    \begin{pmatrix}
              \gamma_{\rm ice} \\  \gamma_2
           \end{pmatrix}
           \right) \ ,
         \ee
         where we have expressed the result in terms of the vector
         \be
         \bld{\tau_V}\equiv 
          \begin{pmatrix}
           1 \\
           {h_2\over h\R}\left( \kappa_{2V}\left(\kappa_{11}+\R\right)-\kappa_{12}\right)
           \end{pmatrix}
         \ee
         and the
          matrix 
         \be
         \bld{\tau}\equiv  
           \begin{pmatrix}
h\kappa_{11} \ \ \ \  & h_2\kappa_{12}
\\
h_2\kappa_{12}  \ \ \ \ 
      & 
      {h_2^2\over h\R}\left(\kappa_{22} (\kappa_{11}+\R)-\kappa_{12}^2\right)
          \end{pmatrix} \ ,
         \ee
         and $\hat{T}$ is defined in equation (\ref{eq:hatt0}). 
         
          The equations of motion for the shell and core librations are
         \be
         {d^2\over dt^2}  \begin{pmatrix}
           C_{\rm ice}{\gamma}_{\rm ice} \\
          C_2\gamma_2
           \end{pmatrix} = \begin{pmatrix}
           T \\
          T_2
          \end{pmatrix}
          \label{eq:eomfull}
                   \ee
     where 
        \be
                 C_2={2\over 5}m_cR_c^2  
         \ee
         is the core's moment of inertia, in terms of the core mass $m_c={4\pi\over 3}\rho_c R_c^3$. 
          In the top panel of Figure \ref{fig:lib} in the body of the paper, we plot the 
          eigenfrequencies of the two free modes from equation (\ref{eq:eomfull}) for Enceladus.
         

\section{Numerical Values for Enceladus}
\label{sec:numbers}

\begin{table}[h!]
\centering
\begin{tabular}{|l|c|c|c|}
\hline
 & \textbf{Symbol} & \textbf{Value} & \textbf{See Eq.} \\
\hline
orbital period
& & 1.3702 day &  \\
\hline
mean motion
& $n$ & $2\pi/$(orbital period)& \\
\hline
eccentricity & $e$ & 0.0047 [current] & \\
\hline
mass  &  &$1.08\times 10^{23}$ g & \\
\hline
radius & $R$    &  252 km   &      \\
\hline
core radius & $R_c$    & $192$ km    &    \\
\hline
shell thickness & $d$    & $\sim 25$ km [current]    &    \\
\hline
water or ice density& $\rho_w$    & 0.93 g/cm$^3$    &      \\
\hline
core density &  $\rho_c$   &  2.47 g/cm$^3$   &      \\
\hline
avg. density &  $\bar{\rho}$   &  1.61 g/cm$^3$   &      \\
\hline
surface gravity& $g$     & $11.35$ cm/s$^2$    &    \\
\hline
shell rigidity& $\mu$     & 4 GPa    &    \\
\hline
Poisson ratio& $\nu$     & 1/3   &    \\
\hline
 & $\kappa_{11}$    &  0.654   &  (\ref{eq:k11})  \\
\hline
&   $\kappa_{22}$  &  0.703    &   (\ref{eq:k22})   \\
\hline
&  $\kappa_{12}$   & -0.194    &  (\ref{eq:k12})  \\
\hline
&   $\kappa_{2V}$  &  0.558    &  (\ref{eq:k2V})   \\
\hline
$h$-Love num.&   $h$  &  1.92   & (\ref{eq:heq}) \& (\ref{eq:heq2})   \\
\hline
& $h_2$    &  1.32   &   (\ref{eq:heq}) \& (\ref{eq:heq2})   \\
\hline
\end{tabular}
\label{tab:numbers}

\end{table}
Numerical values for Enceladus are listed in the table. 
The shell's hardness parameter is then
\be
\R = {d\over 0.83{\rm km}} \label{eq:calrnum}
\ee
(eq. \ref{eq:hardness}). 
And its libration frequency is
\be
\omega_{\rm lib}  &=& n  {2.50\over\sqrt{1+ 1.83d/{\rm  km}}} 
\ee
(eq. \ref{eq:omlib}).

\bibliographystyle{apj}
\bibliography{encel}

\begin{thebibliography}{10}
\expandafter\ifx\csname natexlab\endcsname\relax\def\natexlab#1{#1}\fi

\bibitem[{{Beuthe}(2019)}]{2019Icar..332...66B}
{Beuthe}, M. 2019, \icarus, 332, 66

\bibitem[{{Goldreich} \& {Mitchell}(2010)}]{2010Icar..209..631G}
{Goldreich}, P.~M. \& {Mitchell}, J.~L. 2010, \icarus, 209, 631

\bibitem[{Landau \& Lifshitz(1986)}]{Landau:1986aog}
Landau, L.~D. \& Lifshitz, E.~M. 1986, Course of Theoretical Physics, Vol.~7,
  {Theory of Elasticity} (New York: Elsevier Butterworth-Heinemann)

\bibitem[{Love(1944)}]{love}
Love, A. E.~H. 1944, {A Treatise on the Mathematical Theory of Elasticity}, 4th
  edn. (New York: Dover Publications)

\bibitem[{{Matsuyama} \& {Nimmo}(2008)}]{2008Icar..195..459M}
{Matsuyama}, I. \& {Nimmo}, F. 2008, \icarus, 195, 459

\bibitem[{{Murray} \& {Dermott}(1999)}]{1999ssd..book.....M}
{Murray}, C.~D. \& {Dermott}, S.~F. 1999, {Solar System Dynamics}

\bibitem[{{Shao} \& {Nimmo}(2022)}]{2022Icar..37314769S}
{Shao}, W.~D. \& {Nimmo}, F. 2022, \icarus, 373, 114769

\bibitem[{{Van Hoolst} {et~al.}(2013){Van Hoolst}, {Baland}, \&
  {Trinh}}]{2013Icar..226..299V}
{Van Hoolst}, T., {Baland}, R.-M., \& {Trinh}, A. 2013, \icarus, 226, 299

\bibitem[{{van Hoolst} {et~al.}(2016){van Hoolst}, {Baland}, \&
  {Trinh}}]{2016Icar..277..311V}
{van Hoolst}, T., {Baland}, R.-M., \& {Trinh}, A. 2016, \icarus, 277, 311

\bibitem[{{Vening Meinesz}(1947)}]{1947TrAGU..28....1M}
{Vening Meinesz}, F.~A. 1947, Transactions, American Geophysical Union, 28, 1

\end{thebibliography}

    \end{document}